# Scaling law of the disjoining pressure reveals two – dimensional structure of polymeric fluids

Armando Gama Goicochea[†], and Elías Pérez

Instituto de Física, Universidad Autónoma de San Luis Potosí

Álvaro Obregón 64, 78000 San Luis Potosí, Mexico

## Abstract

A scaling relation for the disjoining pressure of strongly confined polymer fluids is proposed for the first time, which yields directly the scaling exponent, $\nu$, of the radius of gyration for polymers. To test the proposed scaling relation we performed extensive particle – based, coarse – grained computer simulations of polymers confined under $\theta$ – solvent conditions and found that the value of $\nu$ agrees with the expected value for strictly two dimensional chains, $\nu = 4/7$, which points towards the essential correctness of our scaling relation. New approaches are suggested for experimental tests of this scaling law. This work opens up the way to look for other scaling exponents that may reveal new physical regimes and it constitutes an efficient route to determine $\nu$ because the scaling exponent can be obtained with chains of a single polymerization degree by simply reducing the distance between the confining plates.

---

[†] Corresponding author. Email: agama@alumni.stanford.edu

Polymers are known to have scaling properties that depend solely on their degree of polymerization ($N$) and the dimensionality ($d$) of the system [1]. One of such properties is the radius of gyration, $R_G$, which scales as $R_G \sim N^\nu$, with $\nu$ being the scaling exponent. The osmotic pressure ($\pi$) of polymers in solution is another, and has been shown to obey the ideal gas law for dilute concentrations ($c$), while for the semi dilute regime $\pi$ can still scale as an ideal gas of "blobs" whose size is $\xi$ so that $\pi/k_B T \sim \xi^{-d}$ [2, 3]. The transition from dilute to semi − dilute regimes occurs when the chains begin to overlap, at a characteristic concentration $c^*$ that can be obtained from the value of the interchain concentration that equals the intrachain concentration, i. e., when $\xi = R_G$ and $c = c^*$. For values of $c$ above $c^*$, the blob size has to be independent of $N$, which means that $\xi \sim R_G (c/c^*)^y$, where $y = \nu/(1 - \nu d)$. The resulting form of the scaling law for the osmotic pressure is $\pi/k_B T \sim c^{\nu d/(\nu d - 1)}$, as first derived by des Cloiseaux [2]. Here our aim is to propose and test a scaling law for the *disjoining* pressure, $\Pi$, of quasi − two dimensionally (2$d$) confined polymer chains. $\Pi$ is defined as the difference between the component of the pressure tensor normal to the confining surfaces ($P_N$), which are separated by a distance $h$, and the bulk pressure of the unconfined fluid ($P_B$), i. e., $\Pi(h) = P_N(h) - P_B$, therefore it arises solely from the confinement of the fluid, see Fig. 1 [4]. If the confinement is strong (see rightmost image in Fig. 1), the chains are effectively restricted to move on a 2$d$ space.

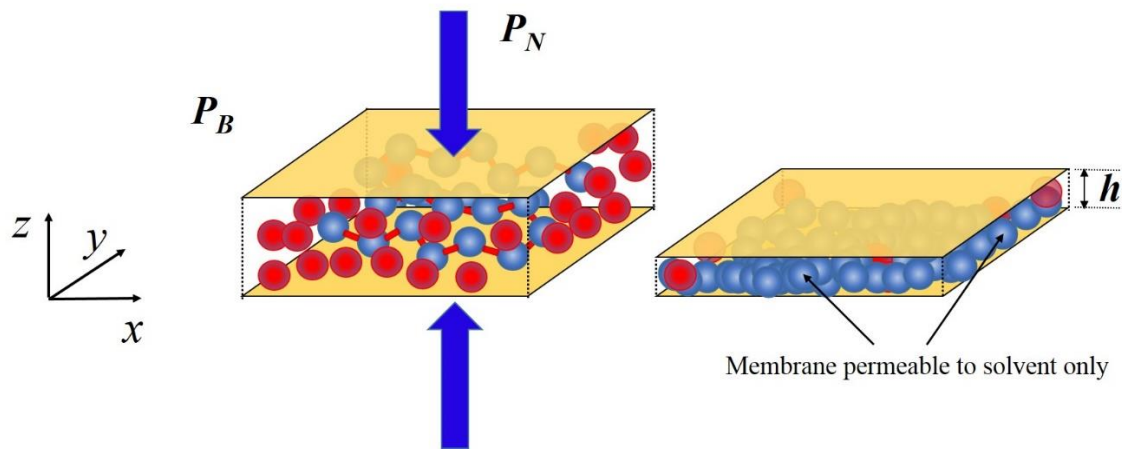

**Figure 1**. (Color online) Schematics of the setup used in the simulations reported here. The disjoining pressure ($\Pi$) is obtained from the pressure normal to the $xy$ – interface ($P_N$) minus the bulk pressure of the fluid ($P_B$), while reducing the distance between the surfaces ($h$). The transversal area is a square of side $L$. The sides of the simulation box, on the $xz$ – plane and on the $yz$ – plane constitute a membrane which is permeable to solvent particles only (in red). Polymer chains are shown in blue.

When the surfaces that confine the fluid are very far apart (large $h$), $\Pi$ decays to zero as a consequence of the decaying interaction between the surfaces and the fluid, while at short separations it may be repulsive or attractive, depending the on the nature of the fluid and the surfaces. This property is of paramount importance in research on colloidal stability and related properties of confined soft matter, because a large, positive $\Pi$ indicates repulsion between surfaces, while negative $\Pi$ represents attraction between them, which may lead to thermodynamic instability [4, 5]. Although $\pi$ and $\Pi$ share qualitatively similar trends, they are different in some important aspects. For example, while $\pi$ is the result of a difference in solute concentration, $\Pi$ is solely due to the confinement of the fluid. Also, $\Pi$ typically shows oscillations that arise from changes in the ordering of molecules in layers parallel to the confining planes, which do not appear in general in $\pi$. We shall study relatively high compression regimes (small $h$ in Fig. 1) which means $d \approx 2$, therefore we propose the following scaling law for the disjoining pressure:

$$\frac{\Pi}{k_B T} \sim \frac{1}{h^* \xi^2} \sim c^y \tag{1}$$

with

$$y = 2\nu/(2\nu - 1), \tag{2}$$

where $\xi$ is the correlation length in $2d$ and $h^*$ is the separation between the surfaces when the polymers start to come into contact laterally as a consequence of the confinement and it is smaller than the lateral length of the surfaces, $h^* < L$. The scaling form of $\xi$ with the concentration is obtained from the following requirements: $\xi$ must be equal to the radius of gyration in $2d$, $R_{G2}$, at $h^*$, and $\xi = R_{G2}(c/c^*)^x$ for $h < h^*$. Here, $c^* = N/(h^* R_G^2)$, and $\xi$ must be independent of $N$ for $h < h^*$ because $\Pi$ should not depend on $N$ in the semi dilute regime. The value of the exponent $x$ is therefore found to be equal to $\frac{\nu}{1-2\nu}$, which yields eq. 1. Notice that $c$ in eq. 1 is defined in $3d$ so eq. 1, while being symbolically equivalent to the $2d$ scaling of the osmotic pressure [1], is fundamentally different from it because the latter requires that $c$ is truly a $2d$ concentration. To our knowledge this is the first time such scaling law is proposed for $\Pi$. It should be remarked that eq. 1 with the exponent $y$ given by eq. 2 is expected to be valid only when $h < h^*$, since it is only when polymers are subjected to strong compression that they move on a quasi - $2d$ space.

To test eq. 1 we perform extensive particle – based, coarse grained Monte Carlo (MC) simulations of polymer chains in a solvent confined by structureless surfaces, in the grand canonical ensemble (fixed chemical potential, $\mu$, volume $V$, and temperature $T$) [6]. Only the solvent monomers are exchanged with the virtual bulk, keeping the total number of polymer chains constant, which is equivalent to having a membrane surrounding the space between the surfaces (see Fig. 1) that is permeable to the solvent only. To sweep through various concentrations in the semi dilute regime we change the distance between the parallel surfaces, $h$, see Fig. 1. By reducing $h$ so that $h < R_G$, the motion of the polymers occurs effectively in $2d$ rather than $3d$. The interaction model we use is the mesoscopic dissipative particle dynamics (DPD) [7]. Details of the method as well as some of its most recent applications can be found elsewhere [8], therefore we outline only what is pertinent here. One of forces that make up the DPD model is the conservative force, given by $\vec{F}_C(r_{ij}) = a_{ij}(1 - r_{ij}/r_c)\hat{e}_{ij}$, where $a_{ij}$ is the strength of the force, which carries all the thermodynamic information of the fluid and is responsible for the excluded volume interactions; $r_{ij}$ is the magnitude of the relative vector between particles $i$ and $j$; $\hat{e}_{ij}$ is a unit vector between them, and $r_c$ is a cutoff distance. The other two DPD forces are the dissipative and random forces whose balance keeps the temperature fixed, as a consequence of the fluctuation – dissipation theorem [9]; all forces conserve momentum and vanish when $r_{ij} > r_c$. The confinement is imposed through an additional, linearly – decaying effective force placed at the ends of the simulation box in the $z$-direction, $F_{wall}(z_i) = a_w(1 - z_i/z_c)$, which is zero if the distance in the $z$ – direction of the particle $i$ with respect to the walls, $z_i$, is larger than $z_c$; the strength of the confining walls' force is given by $a_w$. These featureless walls are meant to be only geometric constraints that keep the polymer chains confined to move in a slice of space where they only have effectively $2d$ degrees of freedom. The $P_N$ component of the pressure tensor is calculated using the virial route [10, 11], and $P_B$ is obtained when the fluid is not confined. Periodic boundary conditions are imposed in the $xy$ – plane, where the system is not bound, but not on the $z$-axis, since the walls are impenetrable. Polymer molecules are constructed by joining DPD beads using freely – rotating harmonic springs [12]; chains of various lengths ranging from $N = 28$ up to $N = 80$ were modeled. One of the advantages of carrying out DPD simulations is that they are reasonably insensitive to finite size effects because of the short range nature of the interactions, which allows one to make correct predictions with relatively small systems [13, 14] in short periods of time. All details of the simulations along with additional data can be found in the Supplementary Information (SI).

Finding the exponent *y* in eq. 1 is equivalent to obtaining the exponent of the radius of gyration, *ν*, as shown in eq. 2. The latter depends on *d* and also on the quality of the solvent; for example, for polymers in good solvent Flory found $\nu = 3/(d+2)$ [15, 1]. Renormalization group calculations have yielded $\nu = 0.588 \pm 0.001$ (3*d*) [16]; from conformal invariance the exact value predicted for *ν* in 2*d* is ¾ [17, 18] while matrix-transfer methods give $\nu = 0.7503 \pm 0.0002$ [19], in good agreement with Flory's formula and MC simulations [20]. For polymer chains in poor solvents, the exponent is inversely proportional to *d*, and this has been verified in 2*d* and 3*d* to a reasonable degree by both experiments and calculations [21]. For $\theta-$ solvent conditions in 3*d* the expected value of *ν* is 0.5 [1], while in 2*d* Vilanove and Rondelez found that $\nu = 0.56 \pm 0.01$ for polymer chains at the air − water interface through surface pressure measurements [22]. A 2*d* estimate for *ν* using an algorithm called "infinitely growing self − avoiding walk" yielded $\nu = 0.57 \pm 0.01$ [23], although it was not shown that such a scheme represented polymers in $\theta$ - solvent. Duplantier and Saleur were the first to show that the exact value for *ν* in 2*d* for chains under $\theta-$ conditions was $\nu = 4/7$ using a self − avoiding walk (SAW) on a hexagonal lattice [24]. MC enumerations for SAW's with attractive nearest neighbor interactions on a square lattice led to the location of the $\theta-$ point for linear polymers in 2*d* with the scaling exponent $\nu = 0.570 \pm 0.015$ [25].

Grassberger and Hegger [26] performed MC calculations for single chains near and below the $\theta$ - point finding reasonable agreement with $\nu = 4/7$ but only for a specific Boltzmann weight. However, we are not aware of any simulations using particle – based models for multiple polymer chains in 2$d$ under $\theta$ – solvent conditions, which are important because experiments with polymers are never truly in 2$d$, in contrast to the numerical and theoretical work cited above. In this work the quality of the solvent is fixed through the choice of the conservative force parameter between particles $i$ and $j$, $a_{ij}$. When $a_{ij} < a_{ii}$, we model a good solvent, while a $\theta$ – solvent is obtained if $a_{ij} = a_{ii}$, for $i \neq j$; poor solvent conditions are obtained if $a_{ij} > a_{ii}$ [27]. The dependence of $\Pi$ on the surfaces separation $h$ is illustrated in Fig. 2 for solvents of different quality, which were determined through the relative values of the interaction parameters.

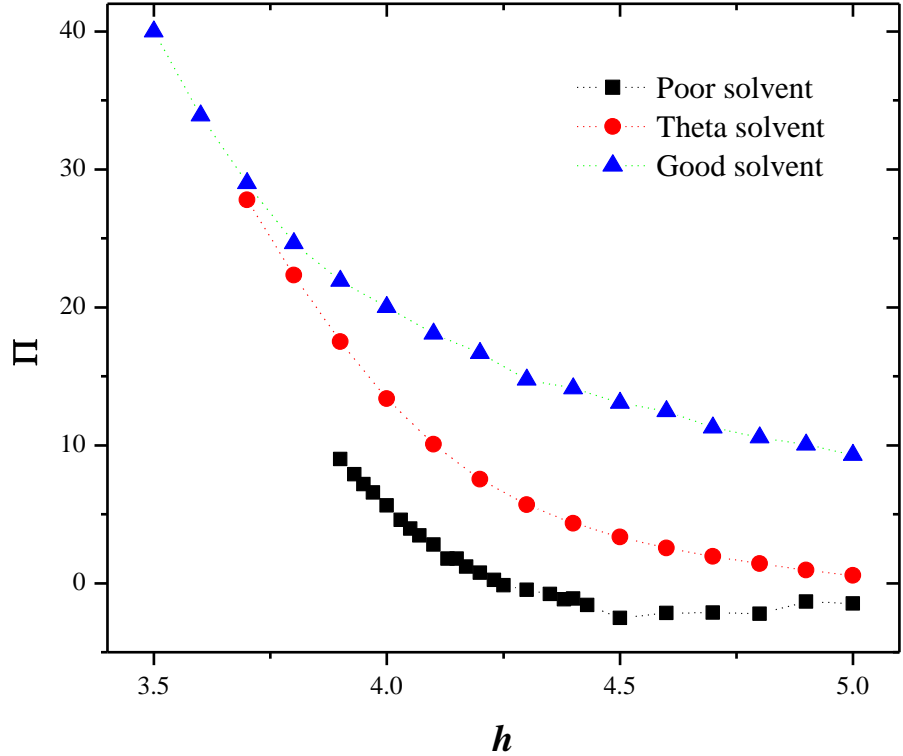

**Figure 2**. (Color online) Dependence of the disjoining pressure ($\Pi$) on the separation between the surfaces, $h$, see also Fig. 1, for different quality of the solvent, defined through the conservative force interaction parameters between particles $i$ and $j$, $a_{ij}$. For good solvent conditions (triangles) we chose $a_{ij} = 65.0$, $a_{ii} = 78.0$; for $\theta-$ solvent $a_{ii} = a_{ij} = 78.0$ (circles), and for poor solvent (squares) $a_{ii} = 78.0$, $a_{ij} = 95.0$. In all cases, there are 10 polymer chains in the simulation box with $N = 28$ each, in addition to the solvent monomers, and the interaction parameter of the surfaces with the fluid particles was chosen as $a_w = 120.0$. All quantities are expressed in reduced DPD units. Lines are only guides for the eye.

The fluid whose disjoining pressure is shown in Fig. 2 is the same for all three cases, and it is made up of 10 polymer chains with $N = 28$ each, plus a fluctuating number of solvent particles, at fixed $\mu$, $V$ and $T$. For poor solvent conditions one notices there is a region where an effective attraction between the walls appears, namely where $\Pi$ is negative, which indicates that a colloidal dispersion under such conditions would be unstable [5]. No hysteresis effects are present here since $\Pi$ is a function of the separation between the walls and the quality of the solvent, although such effects may appear when calculating adsorption isotherms. The strongest repulsion between the surfaces is obtained when the polymer chains are under good solvent conditions, and polymers immersed in a $\theta-$ solvent are found in between the former two cases, as expected. As the compression is increased all cases tend to the same curve, as shown in Fig. 2, which is also to be expected because the number of solvent particles is heavily reduced with the surfaces separation (*h*), leaving a fluid made up mostly of polymer chains, whose interaction with one another is the same ($a_{ii} = 78.0$), regardless of solvent quality. The trends seen in Fig. 2 agree with those found in experiments [28] and in numerical simulations [6, 29].

The results of the MC – DPD simulations [6], performed under $\theta-$ solvent conditions, are shown in Fig. 3. As expected, the disjoining pressure at large compression (large *c*) is found to be independent of the polymerization degree; in fact, we find that $\Pi \sim c^8$ (see the line in Fig. 3), which is the first report of scaling of the disjoining pressure with concentration, to the best of our knowledge. The expected behavior for the *osmotic* pressure of polymer chains in $\theta-$ solvent in $3d$ is $\pi \sim c^3$ [1], which is not supported by the data shown in Fig. 3. Moreover, using eq. 2 and *y* = 8 from Fig. 3 one finds that $\nu = 4/7$ rather than the value expected for $3d$ chains in good solvent, $\nu = 3/5$ [15, 16], in agreement with experiments [30, 31] and confirming the essential correctness of eq. 1.

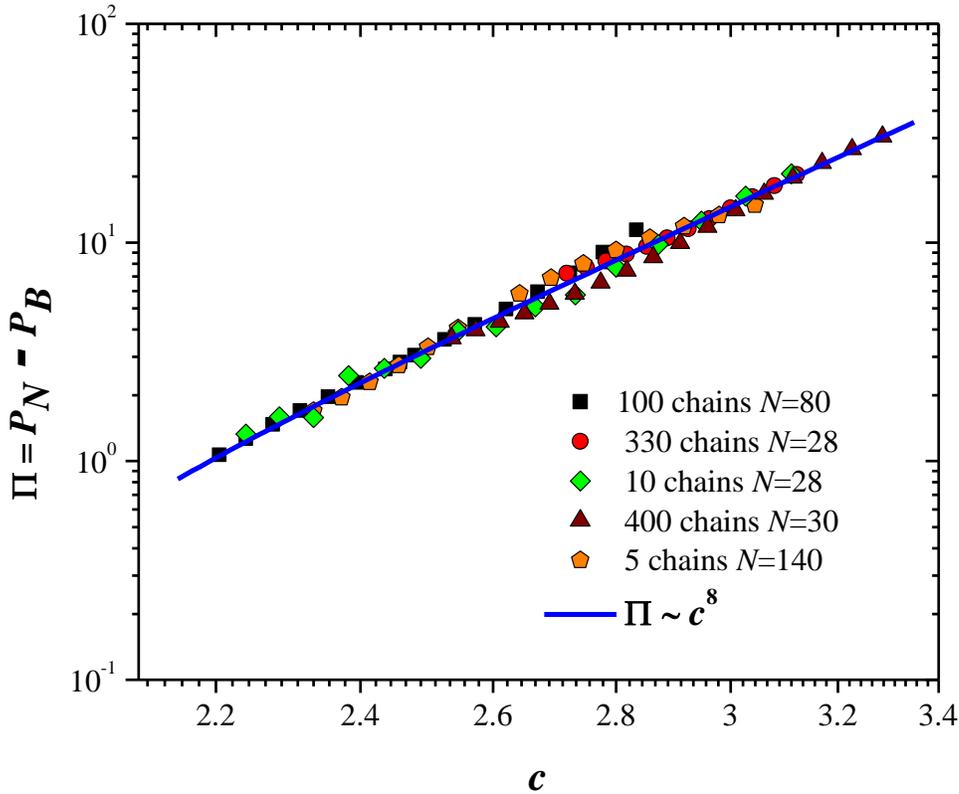

**Figure 3**. (Color online) Dependence of the disjoining pressure ($\Pi$) on the monomer concentration, $c$, for polymer chains of various degrees of polymerization ($N$) under $\theta-$ solvent conditions. The solid line is included as reference and has slope ($y$) equal to 8, which corresponds to a scaling exponent $\nu = 4/7$, see eqs. (1) and (2), and reference [24]. For all cases the surface – monomer interaction was set to $a_w = 120.0$, and the monomer – monomer interaction is $a_{ij} = 78.0$. All quantities are expressed in reduced DPD units.

We have tested the scaling expression in eq. 1 for various values of the polymerization degree, number of chains and size of the simulation box, which for brevity are not shown in Fig. 3 but can be consulted in the SI. The results show that the scaling relation is always fulfilled and the value of the scaling exponent $\nu = 4/7$ is very robust for $\theta-$ solvent conditions. For completeness we performed two additional sets of simulations for polymer chains with $N=28$ under good solvent and poor solvent conditions, finding that $\nu = 0.74 \pm 0.06$ and $\nu = 0.52 \pm 0.04$, respectively, see SI. These predictions are in agreement with the expected $\nu = 0.75$ (good solvent) [17, 18], and $\nu = 0.5$ (poor solvent) [21]. Notice that eq. 2 becomes singular for $\nu = 0.5$, so the poor solvent limit can never truly be achieved. We find that $\nu$ is in good agreement with experiments with large $N-$ polymer chains [22, 30, 31]. The reason why quasi $-2d$ scaling behavior is obtained in these $3d$ systems regardless of solvent quality can be found in the "blob" argument of de Gennes [1]. As the separation between the planes in Fig. 1 is reduced (increasing concentration), the chains begin to overlap and the characteristic length is no longer $R_{G3}$ but instead the size of the blobs, $\xi$, which scales in $2d$ once $h < h^*$. As Fig. 3 shows, $\Pi$ is found to be independent of the polymerization degree, and scales with a $2d$ exponent due to the confinement even though the concentration is calculated in $3d$, and in sharp contrast with the $2d$ scaling of the osmotic pressure which requires a strictly $2d$ concentration [1, 22].

In conclusion, we have shown that the disjoining pressure of strongly confined polymers obeys a simple scaling relation with monomer concentration, $\Pi \sim c^y$. From it we obtain (see eq. 2) the scaling exponent $\nu$ of the radius of gyration of polymers confined by walls at large compression, immersed in a $\theta$- solvent, and find that its value corresponds to the expected $2d$ value $\nu = 4/7$ obtained by other groups using different theoretical and numerical techniques, as well as experiments. The values for good – and poor – solvent conditions are recovered also. Eq. 1 can be tested experimentally using atomic force microscopy by reducing the distance between the cantilever and the sample, in much the same way as the simulations were performed. Polymer films of thickness of the order of 20 Å or less can now be studied where $\Pi$ is measured directly [32]. Microfabricated grooves have been used to measure $\Pi$ on molecularly thin polymer films [33]. These techniques are of paramount relevance not only for the optimization of nanoimprintings [34] and other aspects important to the magnetic recording industry, but also because new scaling regimes can be explored once Π is measured in confined geometries that may lead to understanding of physical phenomena for polymers in reduced dimensions [35]. Lastly, our approach is an improved route to obtain the scaling exponent of the radius of gyration efficiently, one which does not require the simulation of many chains with several values of the polymerization degree, as would be the case for freely moving chains in solution.

AGG would like to thank M. A. Balderas, E. Mayoral and C. Pastorino for educational conversations. This project was funded by CONACYT's grant 242532.

**Keywords**: Disjoining pressure, Radius of gyration, Scaling, Solvent effects, Statistical thermodynamics.

## Table of Contents Text

Polymers strongly confined by parallel surfaces are found to obey two – dimensional scaling laws in their disjoining pressure and radius of gyration, using mesoscopic computer simulations. These scaling characteristics depend on solvent quality as well. Our predictions are in excellent agreement with results obtained for strictly 2D polymers, and can be tested in polymeric fluids using atomic force microscopy, for example.

## Table of Contents Graphic

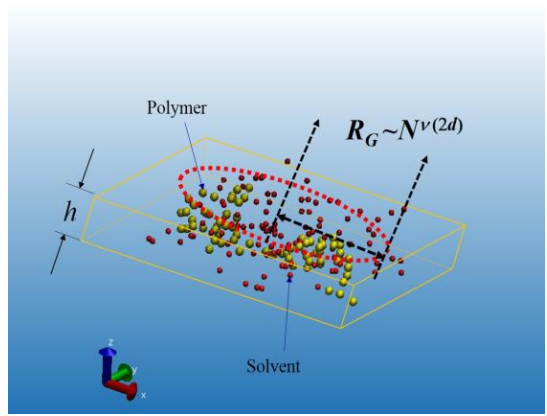